\documentclass[preprint2]{aastex}

\usepackage{graphicx}
\usepackage{amssymb}

\def\deg{\hbox{$^\circ$}}
\def\sun{\hbox{$\odot$}}
\def\earth{\hbox{$\oplus$}}
\def\lesssim{\mathrel{\hbox{\rlap{\hbox{\lower4pt\hbox{$\sim$}}}\hbox{$<$}}}}
\def\gtrsim{\mathrel{\hbox{\rlap{\hbox{\lower4pt\hbox{$\sim$}}}\hbox{$>$}}}}

\def\arcsec{\hbox{$^{\prime\prime}$}}

\received{2011 July 11 17h PDT}

\begin{document}

\title{Measurement of the Electric Current in a Kpc-Scale Jet}

\author{P.P. Kronberg\altaffilmark{1}} \affil{Institute of Geophysics and Planetary Physics, Los Alamos
National Laboratory M.S. B283, Los Alamos NM 87545, and Department of Physics, University of Toronto, 60 St George St., Toronto ON M5S 1A7}

\author{R.V.E. Lovelace\altaffilmark{2}} \affil{Department of Astronomy,
Cornell University, Ithaca, NY 14853-6801}

\author{G. Lapenta\altaffilmark{3}} \affil{Centre for Plasma Astrophysics, Departement Wiskunde, Katholieke Universiteit Leuven, Belgium (EU)}

\author{S.A. Colgate\altaffilmark{4}} \affil{Los Alamos National Laboratory, Los Alamos NM 87545}

\altaffiltext{1}{kronberg@lanl.gov}
\altaffiltext{2}{lovelace@astro.cornell.edu}
\altaffiltext{3}{giovanni.lapenta@kuleuven.be}
\altaffiltext{4}{colgate@lanl.gov}

\begin{abstract}

{We present radio emission, polarization,
and Faraday rotation maps of the radio jet of the galaxy 3C303.
  From this data we derive the
magnetoplasma and electrodynamic parameters of this $50$ kpc long jet. 
     For one component of this jet we  obtain for the first
time a direct determination of a {\it galactic}-scale electric 
 current ($\sim 10^{18}$ A) , and its direction $-$ {\it positive} away from the AGN.
Our analysis strongly supports a model where the jet energy flow is mainly  electromagnetic. }

\end{abstract}


\section{Introduction}

	{ A fundamental conundrum of astrophysical jet models is how energy is
extracted from the accretion flow close to the black hole event horizon.
   The total energy carried by the jets of active galaxies  is estimated to be 
a non-negligible fraction of the supermassive black hole (SMBH) formation energy, $\sim 0.1 M_{\rm bh}c^2$ (Kronberg et al. 2001).   
       The jets  are initially highly relativistic and
low-density and for this reason are thought to be {\it magnetically dominated}
or force-free
with a negligible fraction of the power in particle kinetic energy.}
  That is,
the energy outflow from the accretion disk is
 in the form of a collimated
 `Poynting-flux jet'  as proposed by Lovelace (1976) 
and  Blandford (1976) and subsequently studied in many papers  
(Benford 1978; Lovelace, Wang, \& Sulkanen 1987; Lynden-Bell 1996; Li, et al. 2001; Lovelace, et al. 2002;  2003; Nakamura et al. 2008).
    The model has a current outflow (or inflow) $I_z$ along the spine of the jet of cylindrical radius $r_J$ initially of the order of
 the Schwarzschild radius of the black hole.  
    The associated toroidal magnetic field is responsible for collimating
the jet.
   An equal but opposite ``return current'' flows inward (or outward) at much larger distances from the jet axis so that the net current  outflow from the source is zero. 
      Because the jet current and  its return have opposite signs 
they repel, as a result of their magnetic interaction mediated by the toroidal
magnetic field $B_\phi$.   This repulsion between the jet and
its ({ more spatially distributed}) return current has been demonstrated in  MHD simulations (Ustyugova et al. 2000, 2006, { Nakamura et al. 2008).

        Rotation measure gradients observed { on parsec scales close to the nuclear central black hole have provided evidence consistent with an electric current flow} 
 (Asada {\it et al.} 2002; Gabuzda, Murray, \& Cronin 2004; Zavala \& Taylor 2005).
    { Here, on a scale $10^4$ times larger in a ``mature'' jet, we present a measured estimate of the jet's {net axial} current, and its sign}.

\section{Brief description of the observations}

      { 3C303 was observed in the VLA's most extended ``A'' configuration in Stokes parameters I, Q, U in the 1.4, 4.9 and 15GHz radio bands with maximal
 $u-v$ coverage over an 11.5 hour observing period in April 1981. Calibrated images have been discussed and analyzed in Kronberg (1986) and Lapenta and 
Kronberg (2005 - hereafter LK05). In 2010 and 2011, the 1.4 and 4.9 GHz images were re-edited (flagged), re-calibrated and re-imaged using more 
updated AIPS imaging procedures, partly as a check of the earlier 1980's imaging procedures. Excellent correspondence was found, 
though the more recent procedures permitted better imaging of the faintest structures. Due to the higher noise levels and sparser aperture plane ($u-v$) 
coverage  of the 15 GHz data they are not presented here, though they are useful for confirming the main image features at a resolution of $\sim 0.15 \arcsec$, {and the lack of any 
detectable Faraday rotation between 15 and 4.9 GHz.}}}      

\section{The jet structure and knots of 3C303}

The 3C303 radio source at $z=0.141$ (Figures 1,2) has a one-sided prominent jet with regular repeating knot structures { in  both VLA radio 
(Kronberg, 1976, 1986; Lapenta \& Kronberg 2005 (LK05), and Chandra X-ray bands (Kataoka et al. 2003).
It resides in a relatively sparsely populated intergalactic environment, and at a high galactic latitude 
[($l$,$b$) = +90$^\circ$.5, +57$^\circ$.5].}
Being away from a galaxy cluster, it is well-suited to an analysis of Faraday rotation measures (RM) within
the jet and lobes, since competing Faraday RM from an immediate cluster environment and the Galactic
foreground are small. 
   Given this situation, we applied improved, new determinations of the RM's of neighboring background radio sources 
(Kronberg \& Newton-McGee 2011). { This allows the most accurate available estimate of the ``RM  zero-level'', so that our
observed RM gradients and RM=0 crossings are minimally affected by foreground plasma (Section 5)}.

The above combination of radio and X-ray images plus background source RM's permits the jet's physical parameters, 
including now its current, to be analysed in isolation. 3C303's radiating knots are large, a few { {\it thousand}}
times larger in volume
than those in, for example, the well-studied nearby M87 radio galaxy. The laterally unresolved ``spine'' of its jet 
is surrounded by a ``cocoon'' of relativistic, radio and x-ray-visible, gas mixed with thermal
plasma which has both a low plasma - $\beta$ (ratio of plasma
pressure to magnetic pressure) and a highly ordered magnetic field { structure} (Fig.1).

{ The plasma $\beta$ along with the other plasma parameter, and minimum energy estimates are derived and presented in LK05, 
which can be consulted for further details.
We distinguish the jet spine, of cylindrical radius $r_J$ normal to the propagation direction ($z$) and 
which is laterally unresolved and appears to carry
the jet power, from the ``knots'', which are partially resolved ``cocoons'' around the unresolved ``spine''.}
{ The cocoons contain a mix of magnetized thermal and 
relativistic plasma. We find no detectable growth in the knot widths away from the AGN, 
which puts an {\it upper limit} of 0.7$^{\circ }$ on the opening angle of the jet.} 

The jet's regular structure, combined with the
appearance of a transverse, magnetically coherent, western lobe complex { also} suggests it as a clear 
laboratory of jet disruption. The disruption point is indicated in Figure 1. 
Model simulations by LK05 applied MHD soliton-like solutions of the Grad-Shafranov equation to 
the radio and X-ray radio images of the regular knot structure of the 3C303 jet. 
Their models successfully computed jet stability times and constrained its plasma parameters. 
The { above analysis forms} a basis for this paper, independent of the { MHD} model details in LK05. 
{ It is the visible, synchrotron radiating  cocoon surrounding the jet that enables us to probe the 
current, most of which is probably confined to the unresolved  ``spine''.}
It is also apparent that the jet continues beyond the disruption point in Figure 1, and terminates 
in a bow shock-like extremity to the west. 
To the east, a large polarized loop is also seen, as well as a possibly related ``counter hotspot''
that aligns with the western jet. We do not attempt interepretion of these latter two features
in this {\it Letter}, and defer them to a following paper. 

	Below, we describe a rare opportunity to measure the
Faraday rotation (RM) variation transverse to, and along, a segment of  the jet in knot E3 (Figure 1).
Combined with the full plasma diagnostics, and a 
determination of the RM zero-level from surrounding sources 
this provides, for the first time, a direct estimate of both the magnitude and sign of a jet's
{\it electric current} on kiloparsec scales in a low density intergalactic environment. 

	The partial resolution of the knots in the VLA images at 1.4 and 4.9 GHz 
permits estimates of the relativistic { gas {density} and approximate magnetic field energy} within each 
kpc$-$ scale synchrotron radiating knot (cocoon).

The first stage in measuring the jet current is to calculate the approximate magnetic field strengths in 
the knots using the analysis described in LK05. Note that 
the knots' equipartition magnetic field strength estimates depend only weakly on the volume 
filling $\phi $, as $\phi ^{-2/7}$, and on the upper cutoff frequency of the photon spectrum of the jet's emission. 
The latter extends to 10$^{17}$ Hz (Kataoka {\it et al.}, 2003), and its reduction 
by, say $\times 100$, would reduce the (redshift-corrected)  total energy content 
in the knots by only $30\%$. 
At lower particle energies the uncertain extrapolation of the knots' relativistic particle 
energy spectrum, corresponding to (unobserved) very low radio frequencies, could in principle 
raise our estimate of $|{B}|$ by a modest factor less than unity, however no anomalously steep spectrum between 
1.4 and 4.9 GHz is seen in the region of the 3C303 jet.
 
We estimate a field strength in 
3C303's jet knots, where $B_{\rm knot}\sim 0.5$ milligauss {(LK05)}, measured with a $0.35\arcsec$ beam 
and allowing for some out-of-plane component due to the { expected} helical form of the jet's 
magnetic field.

At the 0.35$\arcsec$ resolution of the 4.9GHz VLA image (Figure 2) the projected line-of-sight 
magnetic field orientation within each knot is parallel to the jet axis. The 4.9GHz
percentage polarizations of knots E$_{1}$, E$_{2}$, and E$_{3}$ are $\sim 27\%$, $21\%$, and $26\%$, 
respectively, and the knots are elongated in the { projected} jet direction by $\sim 2.2\arcsec$.

\section{The Faraday rotation and thermal electron density}

Figure 3 shows the 2-D RM variation within the brighter parts of
the  entire 3C303 system  derived from the $1.4$ and $4.9$ GHz images at a common $1.5\arcsec$ resolution. Values range between $+5$ and $+25$ rad
m$^{-2}$ over a maximum extent of $\sim 50\arcsec$, including 3C303's extended western lobe. No higher RM's 
are seen anywhere within the source boundaries. One test for a 180$\deg$ ambiguity in $\chi(\lambda)$ is to separately  produce Q and U images at the higher 
frequency of 5GHz, where we can test for ``exchanges'' between Q and U along the jet. None were found, meaning that 
{ no substantially higher RM's} occured in or around the jet. The ambiguity-free RM  of 3C303's integrated emission (from polarization data at 6 frequencies) 
is + 18 $\pm$ 2 rad m$^{-2}$ (Simard-Normandin et al. 1981). This is nicely concordant with the RM range in Figure 3, after the -18 rad m$^{-}2$ correction above for the independent foreground RM estimate in $\S 5$ below.  

The jet knot E3 is resolved
just enough both across and along the jet, to show an RM gradient which is { independently} transverse to the jet axis. 
The varying RM in knot E3 also independently 
confirms the presence of some thermal plasma in the cocoon.
{ The RM gradient, $\nabla RM$ around knot E3 is $\sim 10$ radians m$^{-2}{\rm kpc}^{-1}$.
Because the RM in Fig. 3 is resolved in 2D over 10 - 12 independent sampling points in {a projected}
$y - z$ plane around knot E3, we are able to specify both the magnitude and direction 
of $\nabla RM$ within the knot E3 cocoon.}   

Having measured a differential RM over the knot,
its dimension, and the approximate magnetic field strength, we can estimate the non-relativistic plasma density within the knot via the following equation. 
$$
{\rm RM}= 4.1\times 10^{5}\left( {\frac{{n_e }}{{{\rm cm}^{ - 3} }}} \right)\left( {\frac{{B_\parallel  }}{{\rm mG}}} \right)\left( {\frac{x}{{500{\rm pc}}}} \right){{\rm rad}\over {\rm m}^2 }
$$
\begin{equation}
\approx 10~{{\rm rad}\over {\rm m}^2 }~,
\end{equation}
where $x$ is the line-of-sight distance through the knot. Inserting $B_{\parallel}$ and $x$ gives $n_{e}\sim 1.4
\times 10^{-5}$cm$^{-3}$, similar to the value obtained in LK05. 

An independent  estimate of  $n_{e}$ can be made from the observed polarization {\it degree} in the knots. 
On the reasonable assumption that the very regular field geometry in knot $E3$ results from a helical geometry, 
knot-internal Faraday dispersion is limited by the high observed degree of polarization $\simeq 25 \%$ at $5$ GHz.
This test implies a back-to-front, knot-internal differential Faraday rotation at 5GHz of $\lesssim 30^{\circ}$, 
consistent with the knot's RM observed above {(LK05)}.
{ It} is consistent with the above $n_{e}$ estimate  
and gives us confidence in the local $n_e$ scaling that is required 
to estimate the jet current - which we discuss next.

       This low value of $n_e$ outside the jet spine points to a magnetically
confined (or dominated) jet (LK05):   Confinement of the jet by hot gas 
at X-ray temperatures is not possible because $p_{\rm gas}=n_ekT_X \ll { \bf B}^2/8\pi$. A common feature of the magnetic jets is that the 
internal pressure of the magnetic field $B_{\rm knot}^2/8\pi$ (and
relativistic particles) is confined by the pinch force of the external toroidal
 magnetic field $B_\parallel$.  The models give $B_{\rm knot} \sim
 B_\parallel$ (LK05).

\section{Estimate of the electric current in the 3C303 jet}

{ An important} final step in relating ($\nabla $RM) to { an axial} current is to calibrate the RM zero level. 
This was done by averaging discrete source RM's { along neighboring lines of sight to} 3C303 from a recent accurate 
RM dataset (Simard-Normandin et al. 1981, Kronberg \& Newton-McGee 2011). The resulting RM zero-level correction, 
$-18~\pm 4 ~$rad m$^{-2}$, has been applied to the RM scale of Figure 3. With this zero-level correction,
we find that the RM in knot E3 changes sign {\it on} the jet axis, within the limit of our uncertainties. 
Furthermore, its gradient ($\nabla $RM = $n_eB_{\phi}$) is found to be perpendicular to the { (independently measured)} axis of 
knot E3 and the jet axis. This corresponds to the expected RM behavior around an electric current  flowing along a jet axis,

The reversal of the RM sign  on the jet axis matches the signature of a current-carrying cylinder, where 
we are measuring the azimuthal component of a { helical} magnetic field  at  distances $\approx y$ on opposite sides of the 
synchrotron radiating cocoon surrounding the unresolved spine of the jet.
  {In terms the measurables, the inferred current, 
$I_z  \propto  (y/n_e )\nabla RM $
is the total current within the cocoon's  radius $y$.
Over this distance we measure  $\nabla$RM  of $\sim$ + 10 rad m$^{-2}$/kpc. }  Thus, within knot {E3}
\begin{equation}
I_{z} \sim 7.7 \times 10^{18} \left( {\frac{{B_\phi}}{\rm{mG}}} \right)\left( {\frac{y}{{0.5{\rm{kpc}}}}} \right){\rm{A}}~,
\end{equation}
where $B_{\phi}$ is the toroidal component of magnetic field at a distance $y$  in the sky plane from the jet ($z$) axis.
The positive sign of the gradient
$\nabla$(RM) 
indicates that a current of $\sim$3.85
$\times$ 10$^{18}$ A, for $B_{\rm equi}$=0.5 mG, is directed {\it away} from the galaxy nucleus. That is, { within $0.5$ kpc of 
the jet axis} there is a net flow of negative charge toward the AGN core within knot E3.  

{Our result for the jet current
 is subject to different corrections and adjustments:
For example, ({1}) if the jet axis is significantly out of the plane of the sky then
the observed RM gradient can arise from a purely
toroidal field, but a helical field would {in this case} shift the 
RM$=0$ line away from the jet's axis, contrary to the present observations.
 ({2}) In another example if the magnetic field of the jet is not symmetric about
its axis the field estimates will be modified. { Such possibilities, especially (1), underline the importance of 
establishing from observations, an optimally foreground-free RM ($y$,$z$) over the source image, as we have done here.}   
   Considering these uncertainties  and the limited angular resolution, our best estimate of the 
axial jet current at knot E3 is 10$^{18.5 \pm0.5}$ amp\`{e}res. This is the first direct estimate of a current in a kpc-scale
 extragalactic jet.}

\section{Discussion and Implications}

   A magnetically dominated jet
can be modeled as a transmission
line running in the $z-$direction
with the electric potential drop across it $\Delta V$.
This is equal to the potential drop across the BH accretion disk
from its inner- to outer radius (Lovelace 1976;  Lovelace and 
Ruchti 1983). A typical value is $\Delta V \sim 10^{20}{\rm V} (B_{\rm bh}/10^4 {\rm G})
(M_{\rm bh}/10^8 M_\odot)$,  where $B_{\rm bh}$ is the poloidal magnetic field strength 
near the black hole and $M_{\rm bh}$ is its mass  (Lovelace 1976).  
The current flow in { this} ``transmission line''
is  $I_z = \Delta V/Z$,
 where the impedance of a {\it relativistic } jet is
 $
 Z  \sim  c^{-1} ({\rm cgs}) = (4\pi)^{-1}
 ({\mu_0/\epsilon_0})^{1/2}({\rm MKS}) =  30\Omega
 $
(Lovelace 1976).
  More generally, for a jet with bulk axial velocity $u_z$, 
 $
Z= (1/c)(u_z/c)$,
(Lovelace \& Romanova 2003).
The electromagnetic power transported by the jet is  $L_{\rm jet} =I_z^2 Z$ .

We can now apply these ideas  to the 3C303 system.
The total photon luminosity $L_{\rm rad}$
integrated from $10^{8}$Hz to $10^{17}$Hz is about $3.7\times10^{41}$ erg s$^{-1}$ or 3.7$\times$ 10$^{34}$W (LK05).
        The radiated power is expected to be significantly less than the jet power, because the jet power 
goes into a combination of $pdV$ work on the ambient medium, and to energizing electrons and ions 
{ around the jet and} in the outer radiating lobes.
   Thus, we let $L_{\rm rad} = \varepsilon L_{\rm jet}$,
 where $\varepsilon < 1$ is an efficiency factor.
 Substituting the observed values of $L_{\rm rad}$ and
 $I_z$ we find
 \begin{equation}
\varepsilon = {L_{\rm rad} \over I_z^2 Z} \approx
\end{equation}
$$
10^{-3}\left( {\frac{L_{\rm rad}}{{3.7 \times 10^{34} {\rm{W}}}}} \right)\left( \frac{3.85\times 10^{18} {\rm A}}{I_z } \right)^{{{ 2}}}\left({0.1c \over u_z}\right)~,
$$
where we have normalized the jet bulk velocity to $0.1c$.
      Note that  $\varepsilon$ is  safely less than unity, 
and the model of Poynting flux transport of jet energy in
3C 303's black hole/accretion disk system appears to be self-consistent.
   
The system evidently needs to incorporate a ``transducer'' that converts
the Poynting energy flux into high energy particles which then produce synchrotron
radiation.  
{ These issues, and the complex lobe structure will be discussed in a subsequent paper.}

\section*{Acknowledgements}

 We  thank Rick Perley, 
Robert Reid, Justin Linford and Greg Taylor for help and advice with the re-calibrated images,  Hui Li for 
discussion of current flow in jets, and  an anonymous referee for
a valuable correction to our estimate of the jet current.
Support is acknowledged from a Natural Sciences and Engineering Research of Canada Discovery Grant (PPK) and from NSF
and NASA (RVEL).

\clearpage
\begin{figure}[p]
\centering
\includegraphics[height=3.in]{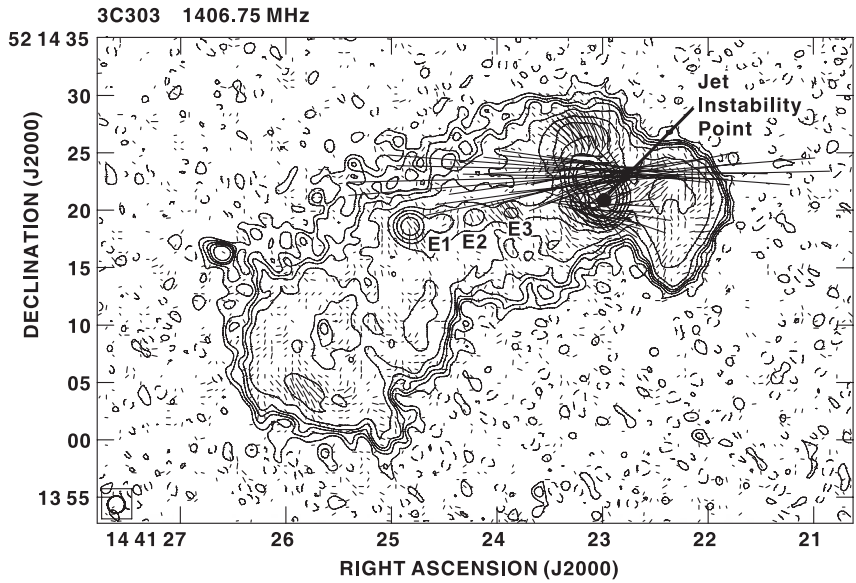}
\caption{1.4 GHz, 1.5$\arcsec$ resolution VLA image of the entire 3C303 system showing an apparent counter-jet knot, the highly polarized
western lobe, and a continuation of the jet to its final western stopping point. Linear polarization lines are scaled such that a 1$\arcsec$ equivalent 
length = 2.5 mJy/beam. I -contours are shown at -0.5, 0.5, 1, 1.5, 2, 3, 6, 12, 24, 50, 100, and 150 mJy/beam. }
\end{figure}
\clearpage
\begin{figure}[p]
\centering
\includegraphics[height=2.in]{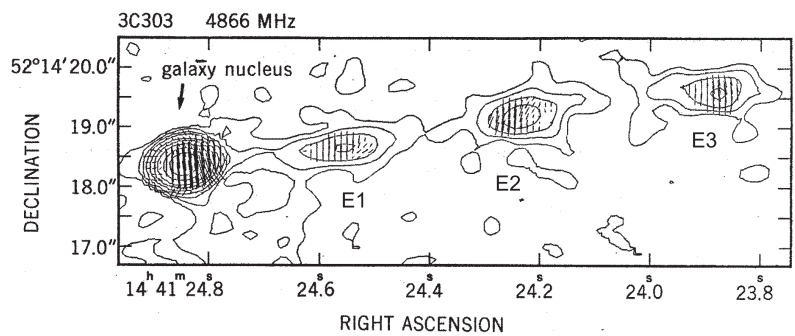}
\caption {A 4.9GHz VLA image of the 
3C303 jet at 0.35$^{\prime\prime}$ resolution showing the 3 prominent, elongated and equally spaced
knots E1, E2, and E3 to the right of the stronger, variable milliarcsec-size galaxy nucleus source.}
\end{figure}
\clearpage
\begin{figure}[]
\centering
\includegraphics[width=5in]{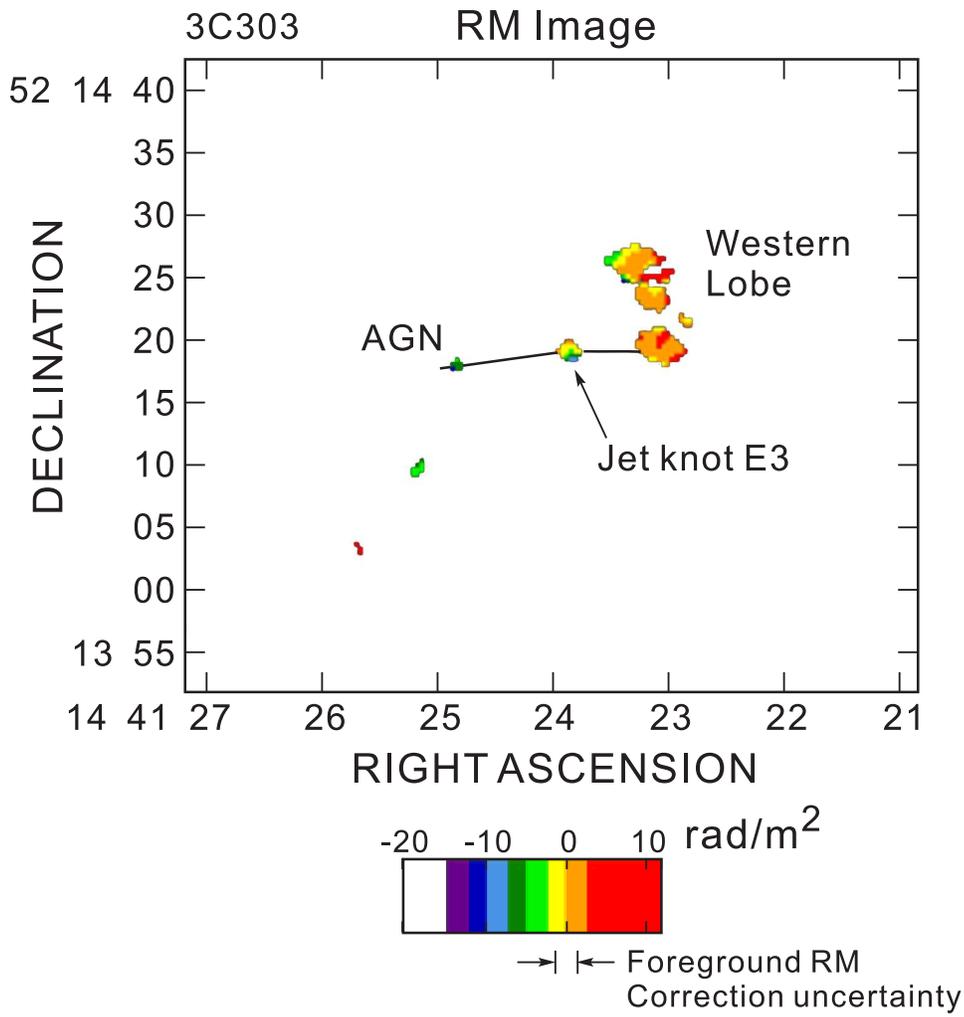}
\caption{Faraday rotation image of the 3C303 radio source at a resolution of 1.5$\arcsec$. The RM zero level has been corrected by 18 $\pm 4$ rad m$^{-2}$ (see~text).}
\end{figure}

\end{document}